\documentclass[prapplied,reprint,showpacs,superscriptaddress]{revtex4-2}

\usepackage{graphicx}
\usepackage{dcolumn}
\usepackage{epsfig}
\usepackage{epsf}
\usepackage{capt-of}
\usepackage{amssymb,amsmath,amsthm,bm,bbm}
\usepackage{empheq,cases}
\usepackage{mathtools}
\usepackage{multirow}
\usepackage[colorlinks=true,linkcolor=blue!50!black,citecolor=red,pdfauthor={ },pdftitle={ },pdfsubject={ },pdfkeywords={ }]{hyperref}
\usepackage{tikz}
\usepackage{pgfplots}
\usepackage{xcolor}
\usepackage{lipsum}
\usepackage{palatino} 
\usepackage[english]{babel}
\usepackage[lmargin=2.1cm, rmargin=2.1cm,tmargin=2.97cm,bmargin=2.97cm]{geometry}
\usepackage{latexsym}
\usepackage{dcolumn}
\usepackage{amsthm} 
\usepackage{amssymb}
\usepackage{amsmath}
\usepackage{graphicx}
\usepackage{multirow}
\usepackage{wrapfig}
\usepackage{subfigure}
\usepackage{bm}
\usepackage{color}
\usepackage{times}
\graphicspath{{Figures/}}

\usepgfplotslibrary{patchplots}
\usetikzlibrary{decorations.text }
\usetikzlibrary{shapes.symbols}
\usetikzlibrary{intersections,shapes.arrows}
\usetikzlibrary{decorations.markings}
\usetikzlibrary{mindmap,trees, backgrounds,fadings,shadows}
\usetikzlibrary{positioning,calc}
\usetikzlibrary{arrows,snakes,shapes}
\usetikzlibrary{patterns}

\definecolor{c1}{RGB}{219,68,56}
\definecolor{c2}{RGB}{74,91,163}

\usepackage{times}

\theoremstyle{definition}

\definecolor{c1}{RGB}{109,63,109} 
\definecolor{c2}{RGB}{243,230,254}

\definecolor{c0}{rgb}{0.27,0.48,0.63}
\definecolor{c1}{rgb}{0.368417, 0.506779, 0.709798}
\definecolor{c2}{rgb}{0.880722, 0.611041, 0.142051}
\definecolor{c3}{rgb}{0.560181, 0.691569, 0.194885}
\definecolor{c4}{rgb}{0.922526, 0.385626, 0.209179}
\definecolor{c5}{rgb}{0.528488, 0.470624, 0.701351}
\definecolor{c6}{rgb}{0.772079, 0.431554, 0.102387}
\definecolor{c7}{rgb}{0.363898, 0.618501, 0.782349}
\definecolor{c8}{rgb}{1, 0.75, 0}
\definecolor{c9}{rgb}{0.647624, 0.37816, 0.614037}
\definecolor{c10}{rgb}{0.571589, 0.586483, 0.}

\usepackage{multirow}
\usepackage{dcolumn}
\newcolumntype{d}[1]{D{.}{.}{#1}}

\begin{document}

\title{Multiple Classical  Noise Mitigation by Multiobjective Robust Quantum Optimal Control}

\author{Bowen Shao}
\affiliation{Shenzhen Institute for Quantum Science and Engineering, Southern University of Science and Technology, Shenzhen, 518055, China}

\author{Xiaodong Yang}
\email{yangxd@sustech.edu.cn}
\affiliation{Shenzhen Institute for Quantum Science and Engineering, Southern University of Science and Technology, Shenzhen, 518055, China}
\affiliation{College of Physics and Optoelectronic Engineering, Shenzhen University, Shenzhen 518060, China}

\author{Ran Liu}
\affiliation{College of Physics and Optoelectronic Engineering, Shenzhen University, Shenzhen 518060, China}

\author{Yue Zhai}
\affiliation{Shenzhen Institute for Quantum Science and Engineering, Southern University of Science and Technology, Shenzhen, 518055, China}

\author{Dawei Lu}
\affiliation{Shenzhen Institute for Quantum Science and Engineering, Southern University of Science and Technology, Shenzhen, 518055, China}
\affiliation{Guangdong Provincial Key Laboratory of Quantum Science and Engineering, Southern University of Science and Technology, Shenzhen, 518055, China}

\author{Tao Xin}
\email{xint@sustech.edu.cn}
\affiliation{Shenzhen Institute for Quantum Science and Engineering, Southern University of Science and Technology, Shenzhen, 518055, China}
\affiliation{Guangdong Provincial Key Laboratory of Quantum Science and Engineering, Southern University of Science and Technology, Shenzhen, 518055, China}

\author{Jun Li}
\email{lij3@sustech.edu.cn}
\affiliation{Shenzhen Institute for Quantum Science and Engineering, Southern University of Science and Technology, Shenzhen, 518055, China}
\affiliation{College of Physics and Optoelectronic Engineering, Shenzhen University, Shenzhen 518060, China}

\begin{abstract}
High-quality control is a fundamental requirement for quantum computation, but practically it is often hampered 
by the presence of various types of noises, which can be static or time-dependent. In many realistic 
scenarios, multiple noise sources coexist, and their resulting noise effects need be corrected to a sufficient order, 
posing significant challenges for the design of effective robust control methods. Here, we explore the method of robust quantum optimal control to generally tackle the problem of resisting multiple noises from a complicated noise environment. 
Specifically, we confine our analysis to unitary noises that can be described by classical noise models. This method employs a gradient-based multiobjective optimization algorithm to maximize the 
control figure of merit, and meanwhile to minimize the  perturbative effects of the noises 
that are allowed for. To verify its effectiveness, we apply this method to a number of examples, including roubust
entangling gate in trapped ion system and robust controlled-$Z$ gate in superconducting qubits, under commonly 
encountered static and time-dependent noises. Our simulation results reveal that robust optimal control 
can find smooth, robust pulses that can simultaneously resist several noises and thus achieve high-fidelity gates.
Therefore, we expect that this method will find wide applications on current noisy quantum computing devices.

\end{abstract}

\maketitle

\section{Introduction} 
For reliable quantum computation, it is a vital task, but a challenging one, to precisely manipulate quantum 
systems in presence of various noises \cite{RevModPhys.88.041001}. 
In reality, noises may originate from a variety of sources, including inaccurate dynamical modeling, unstable 
system parameters, imperfect controls, inevitable interactions with environment, imprecise measurements, etc.  \cite{brif2010control,PhysRevA.91.052306}.
Quantum error correction can generally  suppress the noises to desired level \cite{PhysRevLett.84.2525}.
However, in the noisy intermediate-scale quantum (NISQ) era \cite{preskill2018quantum},  it is still prohibitive 
to realize  quantum error correction, as encoding a single logical qubit needs thousands of physical qubits and 
extensive quantum operations below certain precision thresholds \cite{knill2005quantum,campbell2017roads}. 
Therefore, significant efforts should be made on developing practical strategies, for example, error mitigation 
\cite{PhysRevX.7.021050} and  robust quantum control \cite{PhysRevX.10.031002}, to tackle the noise issue.

Substantial tremendous robust quantum control methods have been proposed in recent decades, such as composite 
pulses \cite{composite1,composite2,CCP13}, dynamical decoupling \cite{dd1,dd2,SAS12}, geometric-formalism-based 
pulse control \cite{PhysRevLett.111.050404,PhysRevLett.125.250403} and sampling-based learning control \cite{dong2015sampling,PhysRevA.89.023402}.
Though these robust control methods have been used in various experiments, their application to general quantum 
engineering tasks still face  a number of challenges: 
(i) Many of the mentioned methods consider the noise effects perturbatively in an analytical manner, and  
thus usually restricted to only the low-order perturbative expansion terms. However,  in situations where achieving 
high-precision target operations is necessary, it is desirable to take the higher-order noise effects into account \cite{PhysRevLett.111.050404,figueiredo2021engineering}.
(ii) Existing methods usually behave well when the noises are quasi-static, but this assumption is not always valid
\cite{oliver2014engineering}. There have been a lot of noise spectroscopy experiments in recent years, revealing 
that time-dependent noises are commonly found in many quantum platforms \cite{PhysRevLett.118.177702,PhysRevApplied.10.044017,Oliver19,PhysRevApplied.15.014033}.
How to resist realistic time-dependent noises with robust control remains currently a difficult task.
(iii) Previously, robust control usually considers resisting one to two types of noises. However, as indicated 
by recent randomized benchmarking experiments, there often exist multiple noises in real quantum devices, which 
must be simultaneously considered for further improvement of control fidelities \cite{PhysRevLett.102.090502,PhysRevA.77.012307,PhysRevA.77.012307}. 
(iv) Realistic noises may have temporal or spatial correlations \cite{PRXQuantum.1.010305}, which poses an 
extra challenge for current robust control methods. Furthermore, things get more complicated when the system size becomes large.
Therefore, a practical robust control method, which can handle the aforementioned issues towards NISQ applications, is still under urgent demand. 

In this work, we study a general robust quantum optimal control method to fight against multiple sources of noises
simultaneously. Quantum optimal control functions by minimizing a settled cost function subject to adjustable 
control parameters with an appropriate optimization algorithm \cite{brif2010control}. To allow for the noise 
impacts, following Ref. \cite{haas_engineering_2019}, we adopt the concept of directional derivative, which 
quantifies the variation of the time-evolution operator caused by given noises. 
The different orders of directional derivatives can be evaluated via Van Loan integral formula \cite{Vanload}, and we 
use their norms as additional cost functions for consideration of robustness. We shall employ the widely used 
gradient ascent pulse engineering (GRAPE) \cite{grape} technique for pulse optimization, hence we refer to the 
whole algorithm as Van Loan GRAPE. Moreover, we add practical constraints of bounded and smoothly varying control 
amplitudes in optimization. Therefore, our method is essentially to solve a multiobjective optimization problem. 
As demonstration examples, we apply the method to search robust optimal control (ROC) pulses for realizing the 
M{\o}lmer–S{\o}rensen gate in trapped ion system and the controlled-Z gate in superconducting system. Various 
types of noise are considered in these test examples, including both static noises and time-dependent noises.
The simulation results reveal that our method is capable of designing robust pulses under complex noisy circumstances 
for high-quality controls. The outline of this work is given as follows. We first introduce Van Loan GRAPE in 
detail in Sec. \ref{method}. Applications in trapped ions and superconducting qubits are then demonstrated in 
Sec. \ref{ion} and Sec. \ref{super}, respectively. Finally, we give brief conclusions and  discussions in Sec. \ref{conclusion}.

\section{Method}\label{method}
Consider a controlled quantum system governed by the Schr\"{o}dinger equation
\begin{equation}\label{eq1}
	\dot{U}(t)=-i[H_{S}+H_{C}(u(t))]U(t) 
\end{equation}
with $U(0)=\mathbbm{1}$ (identity matrix), where $H_{S}$ is the system Hamiltonian, $H_{C}(t)$ 
represents the control Hamiltonian, and $u(t)$ denotes the time-dependent controls. 
 The solution of the above equation at time $t=T$ can be formally written as
\begin{equation}
	U(T)=\mathcal{T} \exp\left(-i\int_0^T dt [H_{S}+H_{C}(u(t))]\right), \nonumber
\end{equation}
where $\mathcal{T}$ is the time-ordering operator.
An essential objective in quantum control is the realization of a desired quantum gate denoted as $U_{\text{tar}}$. This task is typically accomplished by quantum optimal control techniques \cite{brif2010control}, which involve the maximization of the following gate fidelity function with adjustable controls, i.e.,
\begin{equation}\label{fitness function 1}
	\Phi_{0}(u)=|\operatorname{Tr}(U(T)\cdot U_{\text{tar}}^\dagger)|^2/d^2,
\end{equation}
where $d$ represents the system dimension.

\subsection{Noise perturbation}
We now consider that there exist multiple noises during control, which result in deviations of the actual system evolution from the ideal noise-free evolution. The total evolution operator $U_\text{tot}(t)$ with noises can be described by
\begin{equation}\label{pevol}
	\dot{U}_{\text{tot}}(t)=-i[H_{S}+H_{C}(t)+H_{V}(t)]U_{\text{tot}}(t),
\end{equation}
where $H_{V}(t)=\sum_{j}\epsilon_{j}(t)E_{j}(t)$ is the perturbed Hamiltonian, $\epsilon_{j}(t)$ represents 
the $j$th noise, and $E_{j}$ denotes the corresponding noise operator. The solution of Eq. (\ref{pevol}) can be formally expressed as 
\begin{equation}
	U_{\text{tot}}(t)=\mathcal{T} \exp\left(-i\int_0^t dt_1 [H_{S}+H_{C}(t_1)+H_{V}(t_1)]\right), \nonumber
\end{equation}
which is, however, often difficult to calculate because the noise $\epsilon_{j}(t)$ may vary with time randomly. 
For each time instance, the overall evolution is a unitary process, allowing us to take the ensemble average of the noises to define the gate fidelity function corresponding to Eq. (\ref{fitness function 1}), i.e., 
\begin{equation}\label{avgnoise}
	\Phi_{\text{noise}}(u)=\langle|\operatorname{Tr}[ U_\text{tot}(T)  U_{\text{tar}}^\dagger]|^2/d^2 \rangle,
\end{equation}
where $\langle \cdot\rangle$ represents the ensemble average. If the noise evolves slowly relative to the timescales of the controls, it can be considered quasi-static., i.e.,  $\epsilon_{j}(t) =\epsilon_{j}$, thus in this case the gate fidelity  can be simplified to 
\begin{equation}\label{noavgnoise}
	\Phi_{\text{noise}}(u)=|\operatorname{Tr}[ U_\text{tot}(T)  U_{\text{tar}}^\dagger]|^2/d^2.
\end{equation}

To analyze the noise effects, we move to the toggling frame, where the evolution propagator is defined as 
\begin{equation}
	\tilde{U}(t)=\mathcal{T} \exp\left(-i\int_0^t dt_1 \tilde{H}_{V}(t_1)  \right)
\end{equation}
with $\tilde{H}_{V}(t)=U^\dagger (t)H_{V}(t)U(t)$. With these definitions, it holds that $U_{\text{tot}}(t)=U(t)\tilde{U}(t)$, which conveniently separates the effects of noise from the overall evolution.
Subsequently, we expand $\tilde{U}(t)$ with Dyson series \cite{dyson} under the condition that $\epsilon_j(t)$ is small enough, thus get
\begin{align} \label{togham}
	&U_{\text{tot}}(t) = U(t)\left[\mathbbm{1} -i \sum_{j} \int_0^t dt_1 \epsilon_{j}(t_1)\tilde{E}_{j}(t_1) + \right. \\ 
	& \left. (-i)^2\sum_{j,k} \int_0^{t}dt_1 \int_0^{t_1} dt_2 \epsilon_{j}(t_1) \epsilon_{k}(t_2)
	 \tilde{E}_{j}(t_1) \tilde{E}_{k}(t_2) +\cdots 
	\right] \nonumber.
\end{align}
Generally speaking, the noise effects can  be mitigated by reducing the above perturbation terms order by order. There have exist several studies that attempt to analytically find robust controls for minimizing the perturbation terms \cite{PhysRevLett.111.050404,PhysRevLett.125.250403}. However, these methods can only consider the lower orders of the perturbation and one to two types of noises due to the complexity of analysis. To practically quantify and minimize the noise effects, we take use of directional derivatives as proposed in Ref. \cite{haas_engineering_2019} in the optimization  process. As the calculations of directional derivatives  differ slightly depending on whether or not the noise is time-dependent, we clarify in two different scenarios as follows.

\subsection{Directional derivatives  for time-independent noise}
If the noise $ \epsilon_{j}$ is constant, i.e., the noise value does not change with time, then  it can be directly separated from the integral expressions in Eq. (\ref{togham}). The directional derivatives are defined as different orders of derivatives of $U_{\text{tot}}(t)$ with respect to $\epsilon_j$ at $\epsilon_j=0$ \cite{haas_engineering_2019}. For example, the first-order directional derivative is 
\begin{equation}
	\mathcal{D}_{U}^{(1)}(E_{j}) \equiv \frac{d U_{\text{tot}}(t)}{d \epsilon_j} \bigg|_{\epsilon_j=0}= -i U(t)\int_0^t dt_1 \tilde{E}_{j}(t_1),
\end{equation}
and the second-order directional derivative reads
\begin{align}
		 \mathcal{D}_{U}^{(2)}(E_{j},E_{k}) 
	 &\equiv \frac{d^2 U_{\text{tot}}(t)}{d \epsilon_j d \epsilon_k} \bigg|_{\epsilon_j,\epsilon_k=0}  \\ \nonumber
	 &=-2 U(t)\int_0^{t}dt_1 \int_0^{t_1} dt_2  \tilde{E}_{j}(t_1) \tilde{E}_{k}(t_2).
\end{align}
The other orders of directional derivatives can be similarly derived.  By utilizing directional derivatives, we can measure the extent to which a system's evolution deviates from its ideal evolution trajectory due to the presence of noise. However, calculating these derivatives is not an easy task, as it involves resource-consuming integral procedures.

To make the computation of the directional derivatives easier, we apply the technique of Van Loan integral formula \cite{haas_engineering_2019,Vanload}. 
We begin by introducing a block matrix $B(t)$ that incorporates $N$ types of noises defined by the corresponding noise operators $E_j(t), j=1,2,...,N$, i.e.,
\begin{equation}\label{block}
B(t)=\left(\begin{array}{ccccc}
H(t) & E_{1}\left(t\right) & 0 & \ldots  & 0 \\
0 & H(t) & E_{2}\left(t\right) & \ldots  & 0 \\
\vdots & \vdots & \vdots & \ddots & \vdots \\
0 & 0 & 0 & \ldots  & E_{N}\left(t\right) \\
0 & 0 & 0 & \ldots  & H(t)
\end{array}\right)
\end{equation}
where $H(t)=H_S+H_C(t)$. Moreover, we assume that it obeys the differential equation $\dot{V}(t)=-i B(t)V(t)$ with $V(0)$ being the identity matrix.
According to the Van Loan integral formula \cite{Vanload}, the solution of this equation at $t=T$ can then be calculated as
\begin{widetext}
\begin{equation}\label{vt}
V(T)=\mathcal{T} \exp \left(-i\int_{0}^{T} B(t)dt \right) =
\left(\begin{array}{ccccc}
U(T) & \mathcal{D}_{U}^{(1)}(E_{1})(T) & \mathcal{D}_{U}^{(2)}(E_{1}, E_{2})(T) & \ldots  & \mathcal{D}_{U}^{(N)}(E_{1},
\ldots, E_{N})(T) \\
0 & U(T) & \mathcal{D}_{U}^{(1)}(E_{2})(T) & \ldots  & \mathcal{D}_{U}^{(N-1)}(E_{2}, \ldots ,E_{N})(T) \\
\vdots & \vdots & \vdots & \ddots  & \vdots \\
0 & 0 & 0 & \ldots   & \mathcal{D}_{U}^{(1)}(E_{N})(T) \\
0 & 0 & 0 & \ldots   & U(T)
\end{array}\right).
\end{equation}
\end{widetext}
Therefore, by expressing the noise-free Hamiltonian $H(t)$ and the noise operators $E_j(t)$ as a block matrix $B(t)$ in the form shown in Eq. (\ref{block}), we can easily obtain the directional derivatives from the off-diagonal terms of the time-evolution operator $V(T)$ in Eq. (\ref{vt}).

\subsection{Directional derivatives  for time-dependent noise}
For time-dependent noise, we routinely assume that each $\epsilon_{j}(t)$ is a stationary Gaussian process with zero mean, i.e., $\langle\epsilon_{j}(t)\rangle=0$. Additionally, we assume that each pair of the noises is mutually independent, namely $\langle\epsilon_{j}(t_1)\epsilon_{k}(t_2)\rangle=0$ when $j\neq k$.  Under these assumptions, we can characterize each time-dependent noise with its  correlation functions \cite{biercuk2011dynamical,green2013arbitrary}. For simplicity, we only consider the autocorrelation function $\langle\epsilon_{j}(t_1)\epsilon_{j}(t_2)\rangle$ and denote $\langle\epsilon_{j}(t_1)\epsilon_{j}(t_2)\rangle \equiv \epsilon_j^2 \langle \tilde \epsilon_{j}(t_1) \tilde \epsilon_{j}(t_2)\rangle$, where $\epsilon_j$ represents the noise strength for the $j$th noise and $\tilde \epsilon_{j}(t)$ is the corresponding normalized time-dependent noise. 
Usually, the noise has a decaying autocorrelation function, and in such cases, we can approximate the autocorrelation function using a linear combination of exponential functions, i.e., $\langle \tilde\epsilon_{j}(t_1) \tilde \epsilon_{j}(t_2)\rangle = \sum_i a_{ji}e^{b_{ji}(t_2-t_1)},0<t_1<t_2<T$. For example, the autocorrelation function of the typical dephasing  noise in nitrogen-vacancy center system is proportional to an exponential function $e^{-\gamma|t|}$ \cite{PhysRevB.85.155204}. With neglecting cross-correlations between the noises, and taking noise ensemble average of Eq. (\ref{togham}), the equation can be simplified as $\langle U_{\text{tot}}\rangle = U(t)\left[\mathbbm{1}- \sum_j\int_0^{t}dt_1 \int_0^{t_1} dt_2   \langle \epsilon_j(t_1) \epsilon_j(t_2) \rangle \tilde{E}_{j}(t_1) \tilde{E}_{j}(t_2)\right]= U(t)\left[\mathbbm{1}-\sum_{j,i} \epsilon_j^2 \int_0^t dt_1 \int_0^{t_1} dt_2 a_{ji}e^{b_{ji}(t_2-t_1)}\tilde{E}_j(t_1)\tilde{E}_j(t_2) \right]$.
Similarly, the  second-order directional derivative can be expressed  by  
\begin{align}
 	 &\mathcal{D}_{U}^{(2)}(\epsilon_j(t)E_{j}, \epsilon_j(t)E_{j}) 
	 \nonumber  \\ 
	 	 & = -2 \sum_{i} U(t)\int_0^t dt_1 \int_0^{t_1} dt_2 a_{ji}e^{b_{ji}(t_2-t_1)}\tilde{E}_j(t_1)\tilde{E}_j(t_2) \nonumber \\
	 & =   \sum_{i} a_{ji} \mathcal{D}^{(2)}_{U}(e^{b_{ji}}E_{j}, e^{-b_{ji}}E_{j}). \label{td_dd}
\end{align}
Now, to   calculate the above directional derivative, we again use the the technique of Van Loan integral formula \cite{haas_engineering_2019,Vanload}. Concretely, we introduce a set of block matrices for each time-dependent noise $\epsilon_j(t)$
\begin{equation}\label{block2}
	C_{ji}(t)= \left(\begin{array}{ccc}
H(t) & e^{b_{ji}t}E_{j}\left(t\right) & 0  \\
0 & H(t)& e^{-b_{ji}t}E_{j}\left(t\right)  \\
0 & 0 & H(t) \\
\end{array}\right), 
\end{equation}
and solve the corresponding differential equation $\dot{V}_{ji}(t) =- i C_{ji}(t) V_{ji}(t)$. According to the Van Loan integral formula, the solution of this equation at $t=T$ is
\begin{equation} 
V_{ji}(T)= \left(\begin{array}{ccc}
 U(T) & * & \mathcal{D}^{(2)}_{U}(e^{b_{ji}}E_{j}, e^{-b_{ji}}E_{j})(T) \\
0 & U(T)& *  \\
0 & 0 & U(T) \\
\end{array}\right), 
\end{equation}
where the element signified by $*$ is not relevant to our calculations and thus is not shown explicitly. 
Therefore, the directional derivative in Eq. (\ref{td_dd}) can be conveniently calculated using the off-diagonal term    in the above equation.

\begin{figure}
	\includegraphics[width=0.49\textwidth]{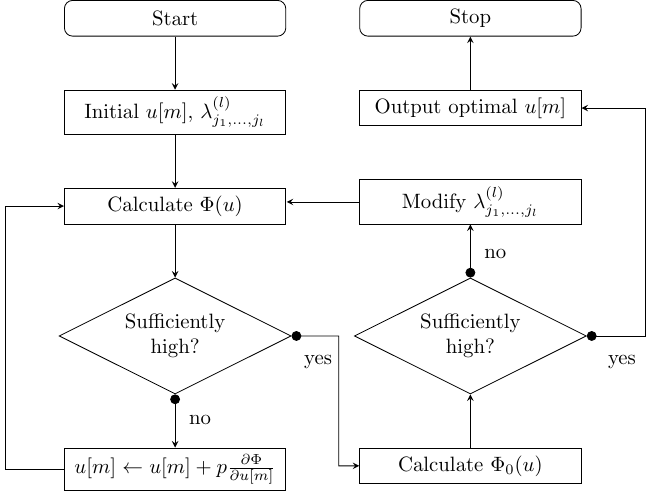}
\caption{Flowchart of Van Loan GRAPE algorithm. The algorithm starts by guessing a random initial control pulse. After a number of iterations, the algorithm produces a   high-fidelity   robust optimal control pulse.}
   \label{flowchart}		
\end{figure}

\subsection{Fitness function and optimization procedure}

From the above descriptions, we clearly see that the directional derivatives can quantitatively characterize the noise perturbative effects of different orders. To find robust controls that can achieve the  control target and meanwhile mitigate the noise effects, we introduce the following multiobjective fitness function 
\begin{equation}
\Phi(u)=\Phi_{0}(u)-\sum_{l} \sum_{j_{1}, \ldots, j_{l}} \lambda_{j_{1}, \ldots, j_{l}}^{(l)}\left\|\mathcal{D}_{U}^{(l)}\left(E_{j_{1}}, \ldots, E_{j_{l}}\right)\right\|^{2},
\end{equation}
where $\Phi_{0}(u)$ is the gate fidelity function introduced above, $\mathcal{D}_{U}^{(l)}\left(E_{j_{1}}, \ldots, E_{j_{l}}\right)$ represents the $l$th-order directional derivative with respect to the noise operators $E_{j_{1}}, \ldots, E_{j_{l}}$, $\lambda^{l}_{j_{1}, \ldots, j_{l}}$ denotes the corresponding weighting coefficient, and $\|\cdot \|$ stands for the Frobenius norm. The weighting coefficients should be set according to the relative significance of the directional derivatives, and are tuned during the optimization procedure. The task then becomes a multiobjective optimization problem. 
 
Generally speaking, there exist various optimization algorithms that can be applied to search optimal controls for maximizing $\Phi(u)$. Here, we choose the famous gradient ascent pulse engineering (GRAPE) algorithm \cite{grape}, which has been widely applied in many quantum engineering tasks \cite{waldherr2014quantum,rong2015experimental,PhysRevLett.118.150503}. 
Combined with the Van Loan integral formula for calculating the directional derivatives, the optimization algorithm we use in this work is called Van Loan GRAPE for short. The flowchart of this algorithm is shown in Fig. \ref{flowchart}. Meanwhile, we briefly describe its algorithm procedures as follows:

\textit{Step 1}. Set the weighting coefficient $\lambda^{l}_{j_{1}, \ldots, j_{l}}$ of the multiobjective fitness function $\Phi(u)$ with appropriate values, randomly initialize the piecewise-constant control fields  $u=(u[m]),m=1,\ldots,M$.

\textit{Step 2}. Evaluate $\Phi(u)$, if not sufficiently high, iteratively do the following steps:

(1) Calculate the gradients of $\Phi(u)$ with respect to $u$, namely $\partial \Phi / \partial u[m]$. 

(2) Update the control fields with  $u[m]\leftarrow u[m]+p\partial \Phi / \partial u[m]$, where $p$ is a suitably chosen step length.

\textit{Step 3}. Evaluate $\Phi_0(u)$, if not sufficiently high, then modify $\lambda^{l}_{j_{1}, \ldots, j_{l}}$  and go to \textit{Step 2}.

\textit{Step 4}. Output the optimal control fields $u$.

In addition, we add the following two constraints during the optimization, so as to meet the requirements of practical applications: (1) bounded control amplitude by setting $|u[m]|\leq \Omega_{\max}, m=1,2,...,M$; (2) smooth control amplitude by requiring the amplitude difference between two adjacent segments to be small.

\section{Robust M{\o}lmer-S{\o}rensen  Gate on Trapped Ions}\label{ion}

Trapped ions are considered to be one of the most promising qubit platforms for quantum computation \cite{Blatt08}. This is due to their long coherence times compared to the operation time, as well as their ability to achieve high-quality initialization and readout \cite{coherence2,high-fidelity}. 
High-fidelity two-qubit entangling gates are essential for universal quantum computation, thus have been extensively studied in trapped ions \cite{PhysRevLett.74.4091,schmidt2003realization,leibfried2003experimental,PhysRevLett.82.1971}. However, various noises, such as Rabi errors and motional heating, can substantially reduce the gate fidelities. To tackle this issue, there have been a number of works focused on developing robust two-qubit gates, but mostly are
designed for one to two types of noises and are not easy to
analytically derive. \cite{robustness2,robustness3,PhysRevLett.121.180502,PhysRevLett.114.120502,PhysRevLett.121.180501}.  
Here, we demonstrate that using the robust quantum optimal control method, high-fidelity two qubit gates on ion traps can be realized under frequently encountered noises

Consider two ions interacting with a laser field of frequency $\omega_l$. The system's Hamiltonian is described by the equation $H_{S}=\omega_{0}\sum_{k=1}^2 {\sigma^k_z}/{ 2}+ \sum_{j=1}^2 \nu_j \hat{a}^\dag_j \hat{a}_j$, where $\sigma^k_{x,y,z}$ are the Pauli matrixes for the $k$th qubit, $\nu_j$ is the eigenfrequency of the $j$th collective motional modes and $\hat{a}^\dag_j,\hat{a}_j$ are the corresponding ladder operators. For convenience, we transform to the ions' resonant rotating frame defined by $e^{-it {\omega_0\sum_{k=1}^2} {\sigma^k_z}/{2}}$, thus the system Hamiltonian becomes
\begin{equation}
	 H_S= \sum_{j=1}^2 \nu_j \hat{a}^\dag_j \hat{a}_j.
\end{equation}
With rotating-wave approximation, the control Hamiltonian can then be expressed as \cite{lee2005phase}
\begin{equation}
	H_C=\frac{1}{2}\sum_{k=1}^2 u_k(t) \sigma_{+}^k e^{i[\sum_{j=1}^2 \eta_{j}(\hat{a}_j+\hat{a}_j^\dag)-\omega t ]} + \text{h.c.}, 
\end{equation}
where $\sigma_+^k =\sigma_x^k+i \sigma_y^k$,  $\omega=\omega_l-\omega_0$, $u_k(t)$ represents Rabi frequency, and $\eta_{j}$ is the Lamb-Dicke parameter. 
Routinely, we restrict to Lamb–Dicke regime \cite{Blatt08}, thus the above equation can be approximated as 
\begin{equation}
		H_C=\sum_{k=1}^2 u_k(t) \cos(\omega t) \left[ \sigma_x^k +\sum_{j=1}^2 \eta_j(\hat{a}_j+\hat{a}_j^\dag) \sigma_y^k \right]. \nonumber
\end{equation}

Suppose that our target is to implement an entangling gate $U_{\text{tar}}   = \exp\left(i {\pi} \sigma^1_y \sigma^2_y/4 \right)$ using the M{\o}lmer-S{\o}rensen  (MS) scheme \cite{PhysRevLett.82.1971,PhysRevLett.82.1835}. 
The original MS gate requires a bichromatic laser field close to one of the motional sidebands, e.g., $\omega_l\approx \omega_0 \pm \nu_1$, and we define $\omega_l-\omega_0 \pm \nu_1 \equiv\pm \delta$.
Moreover, the Rabi frequency is assumed to be constant and not too strong, i.e., $u_1(t)=u_2(t)\equiv \Omega$, $\eta \Omega \ll \delta$ and $\eta_1=\eta_2=\eta$. As such, the system undergoes an adiabatic process, during which the intermediate states will not be populated. By perturbation theory, we obtain an effective Hamiltonian $H_{\text{eff}}=\eta^2\Omega^2/(2\delta) \sigma_y^1\sigma_y^2$. Clearly, if we drive the system for a total time $T_{\text{MS}}=\pi \delta /(2 \eta^2 \Omega^2)$, we can achieve the target MS gate. The drawback of adiabatic
driving is that the gate time is relatively long. A direct way to reduce the gate time is just by decreasing $\delta$ and increasing $\Omega$, and meanwhile make sure that the motional modes are not excited at the end of the evolution. This approach enables us to achieve a relatively fast MS gate, as documented in \cite{PhysRevA.62.022311}; we mark the control waveform obtained in this way as ``original pulse''. Nevertheless, this original MS gate easily suffers from  various noises. We thus use the ROC method introduced above to achieve a robust MS gate through modulating $u_k(t)$.

\subsection{Three types of realistic noises under consideration}

In our investigation, we examine the coexistence of two types of static noises and one time-dependent noise in trapped ions.
The first one is the Rabi-frequency fluctuation \cite{PhysRevLett.114.120502,PhysRevLett.117.060505}, which originates from fluctuations in the laser power or phase, and imperfections in the optics or electronics. 
In general, the bichromatic field consisting of a pair of co-propagating lasers can be subject to distinct noise. For simplicity, we assume the presence of identical noise in both lasers exciting the red and blue sidebands, which can be modeled by $H_{V_1}(t) = \Delta u H_C(t)$. It is worth noting that the Van Loan GRAPE algorithm remains adaptable, allowing for different values of Rabi frequency fluctuations. 
The second one is the frequency fluctuation of the motional modes \cite{PhysRevApplied.13.024022,PhysRevLett.126.250506,PhysRevApplied.19.014014}, which is caused by fluctuations in the trapping potential, couplings to the external environments and variations in the ion's position. It can be descried by  $H_{V_2}(t)=\Delta\omega_1 \hat{a}_1^{\dagger} \hat{a}_1$. 
The third one is the qubit-frequency detuning \cite{PhysRevLett.119.220505,PhysRevLett.121.180502}, which may come from drift of the trapped fields and fluctuations of magnetic fileld in the environment. This noise can be modeled by $H_{V_3}(t)=\epsilon_{\omega_0} (t)E(t)$ with $E(t)=\sum_{k=1}^2 \sigma_z^k/2$. As described above, we assume that $\epsilon_{\omega_0} (t)$ is a stationary Gaussian process with zero mean, and its properties can be characterized by the autocorrelation function $\langle\epsilon_{\omega_0}(t)\epsilon_{\omega_0}(t+\tau)\rangle$. Overall, the perturbed Hamiltonian is $H_{V}(t)=H_{V_1}(t)+H_{V_2}(t)+H_{V_3}(t)$. To search smooth pulses that demonstrate robustness against these noises, we parameterize the control fields and utilize Gaussian smoothing method to ensure the pulses are sufficiently smooth. The method described in Sec. \ref{method} is then applied to optimize the controls $u(t)$ for realizing the robust MS gate. The details about fitness function and gradients are given in the Appendix.
\begin{figure*}
	\includegraphics[width=1\textwidth]{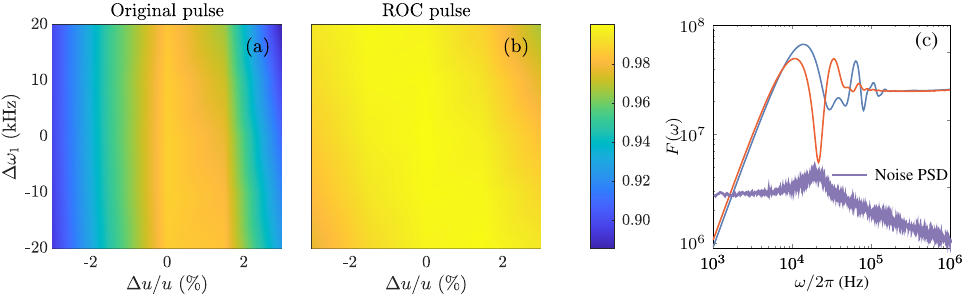}
	\caption{Performance comparison of the original and the robust MS gate under the coexistence of three types of noises. 
  (a) and (b) demonstrate the gate fidelities with respect to the varying Rabi-frequency fluctuation and motional mode frequency fluctuation.
	(c) PSD of the tested noise, and filter functions of the original and the ROC pulse. The time-dependent qubit-frequency detuning is simulated using the parameters $\omega^{\prime}/2\pi = 2 \times 10^4 $ Hz, $A/2\pi = 2 \times 10^4 $ Hz.  All simulations are conducted in a $72$-dimension space where the number of phonons is considered to be 5 and 2, respectively.}
	\label{iontrap2}		
\end{figure*}
\begin{figure}
	\includegraphics[width=0.48\textwidth]{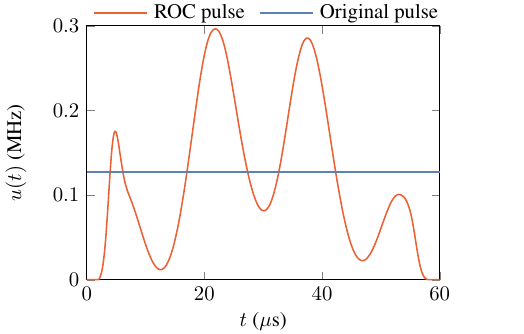}
	\caption{Waveforms of the original and the robust pulse for realizing the MS gate. The original pulse is simulated using parameters $\eta=0.1, \omega=0.983\nu_1$, $\Omega=0.1275$ MHz, $\nu_1=2$ MHz and $\nu_2=2\sqrt{2}$ MHz. The ROC pulse is simulated using parameters $\eta=0.1$, $\omega=0.983\nu_1$, $T=60~\mu$s, $M=600$, $\nu_1=2$ MHz and $\nu_2=2\sqrt{3}$ MHz.}

	\label{iontrap1}		
\end{figure}

\subsection{Simulation results}
The searched ROC pulse, which exhibits a smooth envelope suitable for experimental implementation, is illustrated in Fig. \ref{iontrap1}. We first compare the robustness of the original and the robust pulse under varying Rabi-frequency fluctuations ($-3\%$ to $3\%$) and motional mode frequency fluctuations ($-20$ kHz to $20$ kHz), as shown in Fig. \ref{iontrap2}(a) and Fig. \ref{iontrap2}(b), respectively. It is clear that the robust region for the ROC pulse is much larger than that for the original pulse. This indicates that our ROC pulse can simultaneously resist two types of static noises for improving the gate performance. Moreover, even when $\Delta u/u=0$ and $\Delta \omega_1=0$ Hz, the gate fidelity of the robust MS gate is much higher than that of the original MS gate. This is due to the fact that the modulation of the ROC pulse enhances the control abilities compared to the rectangular original pulse. 

\begin{table}
  {\renewcommand{\arraystretch}{1.5}
  \renewcommand{\tabcolsep}{3pt}
    \begin{tabular}{c|c c c c c}
    \hline
    $\Delta u/u$~($\%$)  &  {$0.5$} &  {$1.0$} &  {$1.5$}  & {$2.0$} & {$2.5$} \\
  
    $\Delta \omega_1~(\text{kHz})$ & {$0.5$}  & {$1.0$} & {$1.5$} & {$2.0$}  & {$2.5$} \\
    $\sqrt{\langle \varepsilon_{\omega_0}^2(t)\rangle}~(\text{kHz}) $ & {$2$}  & {$4$} & {$6$}  & {$8$}  & {$10$}   \\
    \hline
    $\Phi_{\text{noise}}^{\text{original}}$ & {$0.9830$} & {$0.9770$} & {$0.9697$} & {$0.9612$} & {$0.9518$}  \\
    $\Phi_{\text{noise}}^{\text{ROC}}$  & {$0.9999$} & {$0.9999$} & {$0.9997$} & {$0.9995$} & {$0.9993$} \\
    \hline
    \end{tabular}
    \caption{Gate fidelities of the original and the robust MS gate when the tested three types of noises coexist. }
    \label{tabion}
   }
\end{table}

Subsequently, we assess the robustness of the ROC pulse to the considered time-dependent noise.  For convenience, we  move to frequency domain using the filter-function formalism \cite{FF1,FF4}. Specifically, the fitness function in Eq. (\ref{avgnoise}) can be expressed in the following form
 \begin{equation}
 	\Phi_{\text{noise}}(u)=1-\frac{1}{2 \pi} \int_{-\infty}^{\infty}
 \frac{d \omega}{\omega^{2}} S(\omega)F(\omega),
 \label{averagef}
 \end{equation}
 where $S(\omega)$ is the so-called power spectral density (PSD), which is the Fourier transformation of the autocorrelation function $\langle\epsilon_{\omega_0}(t)\epsilon_{\omega_0}(t+\tau)\rangle$ and defined by $S(\omega)=\int_{-\infty}^{\infty}d\tau e^{-i\omega\tau}\langle\epsilon_{\omega_0}(t)\epsilon_{\omega_0}(t+\tau)\rangle$.  
The control filter function $F(\omega)$ can be calculated by 
\begin{equation}
 F(\omega)=  \sum_k\left|-i \omega \int_{0}^{T} d t \operatorname{Tr} \left\{ \mathcal{S}^{\dag} \widetilde{E}_{\text{full}}(t)\mathcal{S}P_k /2^2\right\}  e^{i \omega t}\right|^2, \nonumber
\end{equation}
where  $\{P_k\}_{k=0}^{4^2-1} = \{\mathbbm{1},\sigma_x,\sigma_y,\sigma_z\}^{\otimes 2}$ are Pauli matrices, $\widetilde{E}_{\text{full}}(t) = U^{\dag}(t) (E(t) \otimes \mathbbm{1}_{\nu}) U(t)$ represents the full-space noise operator (toggling frame) with  $\mathbbm{1}_{\nu}$ being the identity operators of  the motional-mode subspaces. The operator $\mathcal{S}$ projects the full-space time-evolution operator into the qubit subspaces. 
 Thus, it provides an effective way to evaluate the quality of a designed control by analyzing the degree of overlap between $S(\omega)$ and $F(\omega)$.
In practice, truncation is necessary to calculate the ladder operators $\hat{a}$ and $\hat{a}^\dag$. We truncate them by considering the phonon number up to five for first motional mode and two for second motional mode in our simulations. 
For simplicity, we consider Lorentzian-type PSD, i.e., $S(\omega) \propto 1/(A^2+(\omega - \omega^{\prime})^2)$  \cite{PhysRevApplied.13.024022,PhysRevLett.126.250506}, as shown in Fig. \ref{iontrap2}(c)  (marked as ``Noise PSD''). The filter functions of the original pulse and the ROC pulse are depicted in Fig. \ref{iontrap2}(c). It's evident that  the filter function  of the ROC pulse has a sharp dip at the characteristic frequency of the noise spectrum, indicating a better performance in resisting this time-dependent noise.

Moreover, we present the gate fidelities of the MS gate under the influence of all three types of noises in Table \ref{tabion}. It is evident that the gate fidelities using the original pulse experience a significant decrease as the noise strength increases. However, with our ROC pulse, the gate fidelities consistently remain above 0.999 in all tested scenarios. This finding highlights the ability of our ROC pulse to effectively resist all three types of noises, thereby enhancing gate performance.

\section{Robust Controlled-Z Gate for Superconducting Circuits}\label{super}

Superconducting circuits are promising for quantum computation due to their ease of fabrication using standard microfabrication techniques, scalability, and ability to achieve long coherence times \cite{superconducting1,superconducting4,kjaergaard2020superconducting,superconducting5}. Two-qubit entangling gates play a crucial role as fundamental components in universal quantum computation and have received significant attention regarding their robustness against various noise sources, including charge noise, magnetic flux noise, and residual couplings \cite{PhysRevApplied.10.054062,PhysRevLett.125.240503,PhysRevLett.123.120502}. Nevertheless, the majority of the existing methods either focus on addressing specific types of noises or encounter challenges when attempting to resist multiple sources of noises simultaneously. Here, we demonstrate that, when several commonly encountered noises coexist, high-fidelity controlled-Z (CZ) gates can be achieved with   robust quantum optimal control.

Consider a system of  two capacitively coupled transmon superconducting qubits, whose Hamiltonian is of the form \cite{superconducting5}
\begin{equation}
	H_{S}= \sum_{i=1}^{2}\left[\omega_i \hat{a}_i^{\dag}\hat{a}_i +\frac{\alpha_i}{2}\hat{a}_i^{\dag}\hat{a}_i^{\dag}\hat{a}_i\hat{a}_i \right] - J(\hat{a}_1-\hat{a}_1^{\dag})(\hat{a}_2-\hat{a}_2^{\dag}), \nonumber
\end{equation}
where $\omega_i$ and $\alpha_i$ are the qubit frequency and the anharmonicity of the $i$th qubit, respectively, and $J$ is the coupling strength.
Usually, two-qubit  entangled gates can be realized through tuning qubit frequencies \cite{PhysRevLett.91.167005,PhysRevLett.123.120502,dicarlo2009demonstration} or applying microwave pulses \cite{PhysRevLett.107.080502,PhysRevB.74.140504}. Here, we aim to realize a two-qubit CZ gate $U_{\text{tar}}=|00\rangle\langle00|+|01\rangle\langle01|+|10\rangle\langle10|-|11\rangle\langle11|$ by adjusting the frequency of the first qubit, thus the control Hamiltonian can be described by
\begin{equation}
	H_C(t) = u(t)\hat{a}_1^{\dag}\hat{a}_1.
\end{equation}
Conventionally, the CZ gate can be achieved using a trapezoid-shaped pulse in combination with auxiliary single-qubit local rotations, i.e., \cite{PhysRevA.87.022309}
 \begin{equation}
 u(t) =  \frac{u_{\text{on}}}{2}\left[\text{erf}\left(\frac{t-t^{\prime}}{\sigma}\right) - \text{erf}\left(\frac{t+t^{\prime}-T}{\sigma}\right)\right],  \label{waveform}
\end{equation}
where $u_{\text{on}}=\omega_2 - \omega_1 -\alpha_2$, erf represents the error function $\text{erf}(x)= \frac{2}{\sqrt{\pi}}\int_{0}^{x} e^{-t^2} dt$, and the pulse switching time is determined by $\sigma$ and $t^{\prime}$. This pulse utilizes the avoided crossing between the energy levels $|11\rangle$ and $|02\rangle$ to realize the CZ gate. However, this original pulse may be sensitive to  certain noises. 
\begin{figure*}[!htb]
	\includegraphics[width=1\textwidth]{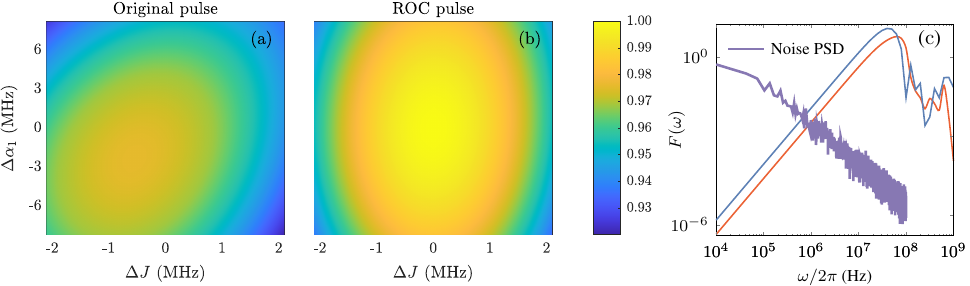}
	\caption{Performance comparison of the original and the robust CZ gate under  the coexistence of three types of noises. 
  (a) and (b) demonstrate the gate fidelities regrading the varying coupling-strength fluctuation and anharmonicity fluctuation.
  (c) PSD of the imported noise, and filter functions of the original and the ROC pulse. The time-dependent  qubit-frequency fluctuation is simulated using  parameters $\omega_l/2\pi = 1\times 10^4$ Hz and $\omega_h/2\pi = 1\times10^8$ Hz.}
	\label{sc2}		
\end{figure*}
	
\begin{figure}
	\includegraphics[width=0.48\textwidth]{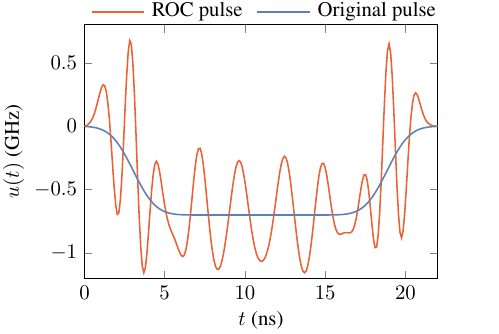}
	\caption{Waveforms of  the original and the ROC pulse for realizing the CZ gate. The parameters for the original pulse are $u_{\text{on}}=-700$ MHz,  $t^{\prime}=62.5$ ns, $\sigma=11.1$ ns. Our ROC pulse is discovered with the parameters $\omega_1 = 7.5$ GHz, $\omega_2 = 6.5$ GHz, $\alpha_1=-300$ MHz, $\alpha_2=-300$ MHz, $J= 25$ MHz, $T=22$ ns and $M=220$.}
	\label{sc1}		
\end{figure}

\subsection{Three types of realistic noises under consideration}	
	
In our investigation, we assume the presence of two types of static noises and one time-dependent noise. 
The first noise of consideration  is  the coupling-strength fluctuation \cite{egger2013optimized,PhysRevLett.125.200503}, which is caused by environmental noises or qubit-frequency drift. This type of noise can be described by $H_{V_1}(t)=\Delta J(\hat{a}_1-\hat{a}_1^{\dag})(\hat{a}_2-\hat{a}_2^{\dag})$. 
The second one is the anharmonicity fluctuation \cite{egger2013optimized}, which arises from defects occurring during the fabrication process. It can be modeled by  $H_{V_2}(t) =  {\Delta \alpha_1}(\hat{a}_1^{\dag}\hat{a}_1^{\dag}\hat{a}_1\hat{a}_1)/2$. 
The third one is the qubit-frequency fluctuation \cite{PhysRevB.77.174509}. It originates from ubiquitous charge or flux noise  and can be modeled by $H_{V_3}(t) = \epsilon_{\omega_1}(t)E(t)$ with $E(t) = \hat{a}_1^{\dagger}\hat{a}_1$, where $\epsilon_{\omega_1}(t)$ is a stationary Gaussian process with zero mean. Therefore, the perturbed Hamiltonian can be summarized as $H_{V}(t) = H_{V_1}(t) + H_{V_2}(t) + H_{V_3}(t)$. We then utilize the method described in Sec. \ref{method} to search for ROC pulses that can generate robust CZ gates. In addition, we also employ the previously introduced Gaussian smoothing technique. More details about fitness function and gradients can be found in the Appendix.

\subsection{Simulation Results}
The obtained ROC pulse exhibits a smooth profile, thus is  suitable for experimental implementations; see Fig. \ref{sc1}. 
Firstly, we compare the robustness of the original and the ROC pulse under the varying  coupling-strength fluctuation and anharmonicity fluctuation, as shown in Fig. \ref{sc2}(a) and \ref{sc2}(b), respectively.
We find that the robust region of the ROC pulse is much larger than that of the original pulse, which reveals the  effectiveness of our method in simultaneously mitigating the two types of static noises. 
Similarly, we use the fitness function Eq. (\ref{averagef}) to evaluate the control performance under the time-dependent noise. The corresponding control filter function can be calculated by  
\begin{equation}
	F(\omega)=  \sum_k\left|-i \omega \int_{0}^{T} d t \operatorname{Tr} \left\{\mathcal{S}^{\dag}\widetilde{E}_{\text{full}}(t) \mathcal{S}P_k /2^2\right\}  e^{i \omega t}\right|^2, \nonumber
   \end{equation}
where $\widetilde{E}_{\text{full}}(t) = U^{\dag}(t) ( E(t) \otimes \mathbbm{1}_{\omega_2}) U(t)$ is the 
full-space noise operator (toggling frame) with  $\mathbbm{1}_{\omega_2}$ being the identity operator of the second-qubit subspace, and  $\mathcal{S}$ projects the full-space time-evolution operator into the two-level qubit subspaces. 
 In our simulations, we truncate the ladder operators $a$ and $a^\dag$ by considering the excited number up to three. 
The noise spectrum is chosen as $1/f$ type \cite{superconducting5} , i.e., $S(\omega) \propto 1/\omega$  $(\omega_l < \omega < \omega_h)$, where $\omega_l$ and $\omega_h$ are the low-frequency and the high-frequency cutoffs, respectively.
Figure \ref{sc2}(c) presents the results of  the original and the robust pulse under the considered time-dependent noise.  It can be seen that the ROC pulse has a lower filter function value than the original pulse across the entire frequency domain, indicating a better noise-filtering ability.
\begin{table}
  {\renewcommand{\arraystretch}{1.5}
  \renewcommand{\tabcolsep}{3pt}
    \begin{tabular}{c|c c c c c }
    \hline
    $\Delta J~(\text{kHz})$   &  $10$ &  $20$  &  $30$  & $40$  & $50$ \\
  
    $\Delta \alpha_1 ~(\text{MHz})$ & $0.1$ & $0.2$  & $0.3$ & $0.4$ & $0.5$  \\
    $\sqrt{\langle \varepsilon_{\omega_1}^2(t)\rangle}~(\text{MHz})  $ & $1$ & $2$ & $3$ & $4$ & $5$  \\
    \hline
    $\Phi_{\text{noise}}^{\text{original}}$ & 0.9734 & 0.9711 & 0.9685 & 0.9658 & 0.9628  \\
    $\Phi_{\text{noise}}^{\text{ROC}}$ & 0.9999 & 0.9997 & 0.9994 & 0.9989 & 0.9982 \\
    \hline
    \end{tabular}
    \caption{Gate fidelities of the original and the robust CZ gate when the tested three types of noises coexist.}
    \label{tabsc}
   }
    \end{table}
Furthermore, we present the results when all three types of noises coexist in Table \ref{tabsc}. It can be seen that the gate fidelities obtained with the original pulse are  below 0.975 in all tested cases. In contrast, the gate fidelities achieved with the ROC pulse are all above 0.998. These findings   indicate that our ROC pulse possesses the capability to effectively mitigate the impact of all the three types of noises, thus leading to considerable enhancement in the overall gate performance.

\section{Discussions and Outlook}\label{conclusion}

Achieving high-quality control is critical for quantum computation, but the presence of various inevitable noises poses a significant challenge. 
Our work demonstrates that robust optimal control offers an effective approach to   devise robust   quantum gates in scenarios where multiple  noises coexist, including static and time-dependent.  Hence it is our  hope that the   method can   find its experimental applications, and not just limited to the examples of ion traps and superconducting circuits here. Also, the method can be easily applied to other quantum control tasks, such as robust quantum sensing \cite{PhysRevX.8.021059} and robust dynamical decoupling \cite{PhysRevApplied.18.054075}.
One important problem is to extend the method   to large-sized systems. As the system size grows, the computational resources required for simulating the system's dynamical evolution will become enormous. Therefore, for future work, we can integrate efficient optimization methods to further
enhance our ability to search for robust pulses \cite{LJ19}, which could be important for NISQ applications.
Additionally, it may be necessary to explore the effects of pulse shaping on nearby qubits from the considered system qubits \cite{PhysRevLett.129.240504}. This exploration could reveal potential gate crosstalk that might degrade the performance of the identified robust pulse in realistic applications.

\section*{Acknowledgments}
We acknowledge support by   the National Natural Science Foundation of China (Grants No. 1212200199,   11975117, 12004165, 12275117, 92065111, and 12204230), Guangdong Basic and Applied Basic Research Foundation (Grant No. 2021B1515020070 and 2022B1515020074),  Guangdong Provincial Key Laboratory (Grant No. 2019B121203002), and Shenzhen Science and Technology Program (Grants No. KQTD20200820113010023, RCBS20200714114820298,  and  RCYX20200714114522109).

%


\setcounter{equation}{0}
\renewcommand\theequation{S.\arabic{equation}} 

\setcounter{figure}{0}
 \renewcommand\thefigure{S\arabic{figure}}


\section*{Appendix}
\subsection{Derivation of Eq. (\ref{averagef}) in the main text: filter-function formalism  for multiqubit system}
For convenience, we consider one time-dependent noise which is described by $H_V(t)=\epsilon(t)E(t)$.
Working in the toggling frame and applying Eq. (\ref{togham}), the gate fidelity in Eq. (\ref{avgnoise}) then becomes 
\begin{equation}
 \Phi_{\text{noise}}(u)\approx  
 1-\frac{1}{d}\int_0^T dt_1 \int_0^{T} dt_2 \langle \epsilon(t_1)\epsilon(t_2) \rangle \text{Tr}[\tilde{E}(t_1)\tilde{E}(t_2)].
\end{equation}
Noting that $\text{Tr}[\tilde{E}(t)]=\text{Tr}[U^{\dag}(t)EU(t)]=\text{Tr}[E]=0$ and $\tilde{E}(t)= \sum_{k=1}^{d^2} X_k(t)P_k$, where $X_k(t)=\text{Tr}[\tilde{E}(t)P_k]/d$ and $\{P_k\}_{k=0}^{d^2-1}=\{\mathbbm{1}_d,\sigma_x,\sigma_y,
\sigma_z\}^{\otimes n}$ are Pauli matrices, the above equation can be simplified to 
\begin{align}
   & \Phi_{\text{noise}}(u)
    =1-\frac{1}{d}\int_0^T dt_1 \int_0^{T} dt_2 \langle\epsilon
	(t_1)\epsilon(t_2)\rangle \text{Tr}[\tilde{E}(t_1)\tilde{E}(t_2)] \nonumber \\
    &= \int_0^T dt_1 \int_0^{T} dt_2 \int_{-\infty}^{\infty} d \omega \frac{S(\omega)}{2\pi} e^{i\omega(t_1-t_2)} 
	\sum_{k=0}^{d^2-1} X_k(t_1)X_k(t_2)\nonumber\\
    &= \frac{1}{2\pi} \sum_{k=0}^{d^2-1} \int_{-\infty}^{\infty} \frac{d\omega}{\omega^2} S(\omega) R_k(\omega)R_k^{*}
	(\omega) \nonumber\\
	&=\frac{1}{2\pi}  \int_{-\infty}^{\infty} \frac{d\omega}{\omega^2} S(\omega) F(\omega).
\end{align}
where $R_k(\omega)= -i\omega\int_0^T dt \text{Tr}[\tilde{E}(t)P_k/d] 
e^{i\omega t}$ and $F(\omega)=\sum_{k=1}^{d^2} |R_k(\omega)|^2$.
The equation for incorporating multiple noise sources can be derived in a similar manner.

\subsection{Fitness function and gradients of the MS gate function}\label{iondetail}
The fitness function of the MS gate can be explicitly expressed as 
\begin{align}
  \Phi(u) &= - \lambda_{1}^{(1)}\left\|\mathcal{D}_{U}^{(1)}\left(H_C\right)\right\|^{2} - \lambda_{2}^{(1)}\left\|\mathcal{D}_{U}^{(1)}\left(\hat{a}_1^{\dag}\hat{a}_1\right)\right\|^{2}  \nonumber \\ 
  & -\lambda_{3}^{(2)}\sum_{i} a_{i}\left\| \mathcal{D}^{(2)}_{U}(e^{b_{i}}E, e^{-b_{i}}E)\right\|^{2} + \Phi_{0}(u) +  \cdots, \nonumber
  \end{align}
  where $E=(\sigma_z^1+\sigma_z^2)/{2}$. The first and second term denote the first-order effects of the two time-independent noises, the third term represents the effect of the time-dependent noise,  and the last one indicates the noise-free gate fidelity.
 The two block matrices shown in Eq. (\ref{block}) and Eq. (\ref{block2}) can be explicitly written as 
   \begin{equation}\label{blocks1}
    B(t)=\left(\begin{array}{ccc}
    H(t) & H_C\left(t\right) & 0  \\
    0 & H(t) & a_1^\dag a_1  \\
    0 & 0 & H(t)  \\
    \end{array}\right)
    \end{equation} 
    and 
    \begin{equation}\label{blocks2}
      C_{ji}(t)= \left(\begin{array}{ccc}
    H(t) & \frac{e^{b_{ji}t}}{2}[\sigma_z^1+\sigma_z^2] & 0  \\
    0 & H(t)& \frac{e^{-b_{ji}t}}{2}[\sigma_z^1+\sigma_z^2]  \\
    0 & 0 & H(t) \\
    \end{array}\right), 
    \end{equation}
  Let the evolution be divided into $M$ slices of equal length $\Delta t = T/M$. Assuming that $\Delta t$ is small, we can consider the control Hamiltonian, the block matrix Eq. (\ref{blocks1}) and Eq. (\ref{blocks2}) to be constant within each slice. Denote the control Hamiltonian and two block matrices at the $m$-th slice as $H_C[m]$, $B[m]$ and $C_{ji}[m]$, respectively. Let $U_m = \exp{\{-i \Delta t(H_S + H_C[m])\}}$ , $V_m = \exp{(-i \Delta tB[m])}$, $V_{ji_m} = \exp{(-i \Delta tC_{ji}[m])}$  the total time evolution operator is then given by $U(T) = U_M \cdots U_1$, $V(T) = V_M \cdots V_1$, $V_{ji}(T) = V_{ji}^{[M]} \cdots V_{ji}^{[1]}$. The VanLoan GRAPE algorithm requires the gradient $g$ of the function $\Phi(u)$ with respect to the control parameters. This gradient can be evaluated according to
  \begin{widetext}
  \begin{align} \label{gradient}
      g[m]  & =  - \lambda_{1}^{(1)} \frac{ \partial\left\|\mathcal{D}_{U}^{(1)}\left(H_C\right)\right\|^{2}}{\partial u[m]} - \lambda_{2}^{(1)} \frac{ \partial\left\|\mathcal{D}_{U}^{(1)}\left(\hat{a}_1^{\dag}\hat{a}_1\right)\right\|^{2}}{\partial u[m]}  -\lambda_{3}^{(2)}\sum_{i} a_{i} \frac{\partial \left\| \mathcal{D}^{(2)}_{U}(e^{b_{i}}E, e^{-b_{i}}E)\right\|^{2}}{\partial u[m]} + \frac{\partial \Phi_0}{\partial u[m]} -\cdots \nonumber \\
      & \approx - 2\lambda_{1}^{(1)} \operatorname{Re}\left\{ \operatorname{Tr}[( V_M \cdots (-i\Delta t\frac{\partial B[m]}{\partial u[m]}) \cdots V_1)\{1,2\} V^{\dag}(T)\{1,2\}] \operatorname{Tr}^{\ast}[V(T)\{1,2\}V^{\dag}(T)\{1,2\}]\right\}  \nonumber \\
      & - 2\lambda_{2}^{(1)} \operatorname{Re}\left\{ \operatorname{Tr}[( V_M \cdots (-i\Delta t\frac{\partial B[m]}{\partial u[m]}) \cdots V_1)\{2,3\} V^{\dag}(T)\{2,3\}] \operatorname{Tr}^{\ast}[V(T)\{2,3\}V^{\dag}(T)\{2,3\}]\right\}  \nonumber \\
      &- 2\lambda_{3}^{(2)}\sum_i a_i \operatorname{Re}\left\{ \operatorname{Tr}[( V_{ji}^{[M]} \cdots (-i\Delta t\frac{\partial C_{ji}[m]}{\partial u[m]}) \cdots V_{ji}^{[1]})\{1,3\} V_{ji}^{\dag}(T)\{1,3\}] \operatorname{Tr}^{\ast}[V_{ji}(T)\{1,3\}V_{ji}^{\dag}(T)\{1,3\}]\right\} \nonumber  \\
     & +  2\operatorname{Re}\left( \operatorname{Tr}[U_M\cdots (-i \Delta t \frac{\partial H_C[m]}{\partial u[m]}) \cdots U_1  U_{\text{tar}}^{\dag} ]\operatorname{Tr}^{\ast}[U(T)U_{\text{tar}}^{\dag}]\right) + \cdots,\\
     \nonumber
  \end{align}
\end{widetext}
where $A\{i,j\}$ represents the block matrix of row $i$ and column $j$ of matrix $A$. For example, in the Eq. (\ref{vt}), $V(T)\{1,2\} = \mathcal{D}_{U}^{(1)}(E_{1})(T)$. More specifically, 
\begin{equation}
  \frac{\partial B[m]}{\partial u[m]}=\left(\begin{array}{ccc}
    \frac{\partial H_C[m]}{\partial u[m]} & \frac{\partial H_C[m]}{\partial u[m]} & 0  \\
  0 & \frac{\partial H_C[m]}{\partial u[m]} & 0  \\
  0 & 0 & \frac{\partial H_C[m]}{\partial u[m]}  \\
  \end{array}\right) ,\nonumber
  \end{equation} 
  \begin{equation}
    \frac{\partial C_{ji}[m]}{\partial u[m]}= \left(\begin{array}{ccc}
      \frac{\partial H_C[m]}{\partial u[m]} & 0 & 0  \\
  0 & \frac{\partial H_C[m]}{\partial u[m]}& 0  \\
  0 & 0 & \frac{\partial H_C[m]}{\partial u[m]} \\
  \end{array}\right), \nonumber
  \end{equation}
 $$ \frac{\partial H_C[m]}{\partial u[m]} =\cos(\omega m \Delta t) \sum_{k=1}^2 \left[\sigma_x^k + \sum_{j=1}^2 \eta_j (\hat{a}_j +\hat{a}_j^\dag)\sigma_y^k\right]. $$ 
To accelerate the convergence speed, we apply the quasi-Newton second-order method \cite{PhysRevA.83.053426,JMR212} in the optimization process. 
This can be conveniently implemented using the built-in function ``fmincon" in Matlab, selecting the limited-memory Broyden-Fletcher-Goldfarb-Shanno  (LBFGS) option.
\subsection{Fitness function and gradients of the CZ gate function}\label{scdetail}
 Here, we consider two time-independent noises and one time-dependent noise. Similarly, for the fitness function, the two block matrices shown in Eq. (\ref{block}) and Eq. (\ref{block2}) transform into
\begin{align}
  \Phi(u) &= \Phi_{0}(u) - \lambda_{1}^{(1)}\left\|\mathcal{D}_{U}^{(1)}\left(\hat{a}_1^{\dag}\hat{a}_1^{\dag}\hat{a}_1\hat{a}_1\right)\right\|^{2} \nonumber \\
  &- \lambda_{2}^{(1)}\left\|\mathcal{D}_{U}^{(1)}\left((\hat{a}_1-\hat{a}_1^{\dag})(\hat{a}_2-\hat{a}_2^{\dag})\right)\right\|^{2} \nonumber \\
  & -\lambda_{3}^{(2)}\sum_{i} a_{i}\left\| \mathcal{D}^{(2)}_{U}(e^{b_{i}}E, e^{-b_{i}}E)\right\|^{2} -  \cdots, \nonumber\\
    B(t)&=\left(\begin{array}{ccc}
    H(t) & \hat{a}_1^{\dag}\hat{a}_1^{\dag}\hat{a}_1\hat{a}_1 & 0  \\
    0 & H(t) & (\hat{a}_1-\hat{a}_1^{\dag})(\hat{a}_2-\hat{a}_2^{\dag})  \\
    0 & 0 & H(t)  \\
    \end{array}\right)  \nonumber
    \end{align} 
    and 
    \begin{equation}\label{blocks4}
      C_{ji}(t)= \left(\begin{array}{ccc}
    H(t) & {e^{b_{ji}t}}\hat{a}_1^{\dagger}\hat{a}_1 & 0  \\
    0 & H(t)& {e^{-b_{ji}t}}\hat{a}_1^{\dagger}\hat{a}_1  \\
    0 & 0 & H(t) \\
    \end{array}\right), \nonumber
    \end{equation}
    respectively, where $E=\hat{a}_1^{\dagger}\hat{a}_1$.
    This gradient can be computed using Eq. (\ref{gradient}) and involves substituting the matrices $B$ and $C_{ji}$ according to the provided expressions. The gradients of $B[m]$, $C_{ji}[m]$ and $H_C[m]$ will also be adjusted using the following  formula:
    \begin{align}
      \frac{\partial B[m]}{\partial u[m]} & =\left(\begin{array}{ccc}
        \frac{\partial H_C[m]}{\partial u[m]} & \frac{\partial H_C[m]}{\partial u[m]} & 0  \\
      0 & \frac{\partial H_C[m]}{\partial u[m]} & 0  \\
      0 & 0 & \frac{\partial H_C[m]}{\partial u[m]}  \\
      \end{array}\right) , \nonumber\\       
        \frac{\partial C_{ji}[m]}{\partial u[m]} & = \left(\begin{array}{ccc}
          \frac{\partial H_C[m]}{\partial u[m]} & 0 & 0  \\
      0 & \frac{\partial H_C[m]}{\partial u[m]}& 0  \\
      0 & 0 & \frac{\partial H_C[m]}{\partial u[m]} \\
      \end{array}\right), \nonumber \\
      \frac{\partial H_C[m]}{\partial u[m]} & =\hat{a}_1^\dag \hat{a}_1. \nonumber
      \end{align}
      
\end{document}